\documentclass[%
 reprint,
 amsmath,amssymb,
 aps,
]{revtex4-2}

\usepackage{graphicx}
\usepackage{dcolumn}
\usepackage{bm}
\usepackage{mathrsfs}
\usepackage{xcolor}

\begin{document}


\title{Nonlinear Optical Joint Transform Correlator for Low Latency Convolution Operations}

\author{Jonathan K. George}
\author{Maria Solyanik-Gorgone}
\affiliation{
 Department of Electrical and Computer Engineering, George Washington University, 800 22nd Street, NW, Washington, DC 20052
}
 
\author{Hangbo Yang}
\author{Chee Wei Wong}
\affiliation{
 Department of Electrical and Computer Engineering, University of California, Los Angeles, 420 Westwood Plaza, Los Angeles, CA 90095
}

\author{Volker J. Sorger}\email{sorger@gwu.edu}
\affiliation{
 Department of Electrical and Computer Engineering, George Washington University, 800 22nd Street, NW, Washington, DC 20052
}
\affiliation{
 Optelligence LLC, 10703 Marlboro Pike, Upper Marlboro, MD 20772, USA
}

\date{\today}

\begin{abstract}
Convolutions are one of the most relevant operations in artificial intelligence (AI) systems. High computational complexity scaling poses significant challenges, especially in fast-responding network-edge AI applications. Fortunately, the Convolution Theorem can be executed \lq on-the-fly\rq\ in the optical domain via a joint transform correlator (JTC) offering to fundamentally reduce the computational complexity. Nonetheless, the iterative two-step process of a classical JTC renders them unpractical. Here we introduce a novel implementation of an optical convolution-processor capable of near-zero latency by utilizing all-optical nonlinearity inside a JTC, thus minimizing electronic signal or conversion delay. Fundamentally we show how this nonlinear auto-correlator enables reducing the high $O(n^4)$ scaling complexity of processing two-dimensional data to only $O(n^2)$. Moreover, this optical JTC processes millions of channels in time-parallel, ideal for large-matrix machine learning tasks. Exemplary utilizing the nonlinear process of four-wave mixing, we show light processing performing a full convolution that is temporally limited only by geometric features of the lens and the nonlinear material's response time. We further discuss that the all-optical nonlinearity exhibits gain in excess of $>10^{3}$ when enhanced by slow-light effects such as epsilon-near-zero. Such novel implementation for a machine learning accelerator featuring low-latency and non-iterative massive data parallelism enabled by fundamental reduced complexity scaling bears significant promise for network-edge, and cloud AI systems.
\end{abstract}


\maketitle

\section{Introduction}

Joint transform correlators (JTCs) are versatile tools that allow full amplitude-and-phase convolution and/or correlation of two signals with latency, fundamentally limited only by the physical performance of the active element in Fourier domain and the speed of light. One of the main functions of JTCs, convolution operation, becomes increasingly popular in AI and ML due to providing spatial context to inputs while simultaneously reducing the network complexity from fully connected to locally connected \cite{fukushima1982neocognitron, aghdam_guide_2017, gu_recent_2018}. These have elevated convolutional operations from their beginnings as simple match filters \cite{kabrisky1970theory} to an integral piece of modern computing hardware \cite{jouppi2017datacenter}, remaining, however, one of the main consumer of computational power in electronic ML hardware. 

On the other hand, the two implementations of JTC's, discussed in this communication, are capable of drastically reducing the scaling complexity while being highly parallelizable in nature. These make them potentially extremely impactful in the field of optical processing and machine vision, edge computing and smart home technologies. The technological market can help overcome outdated challenges of classical JTCs by providing a zoo of modern nonlinear (NL) materials, ultra-fast and high-resolution converters such as cameras, light modulators as well as high-speed electronic interconnect supported by ICs. Supplied a fast enough electronic interconnect, JTC's have a potential to also revolutionize  the ML fields of medical image analysis \cite{litjens2017survey}, speech recognition \cite{cho2014learning}, language translation \cite{graves2013speech}, autonomous driving \cite{bojarski2016end}, inverse design problem solving \cite{tahersima2019deep,liu2018training,sajedian2019finding,qian2020deep}, games \cite{mnih2013playing}, and robots \cite{gu2017deep} to name a few.

Earlier designs for optical correlators often involved a charge-coupled device (CCD) interfaced light-modulator to modulate the probe field with the intensity distribution of the Fourier transformed data and filter signals, see \cite{goodman2005introduction, javidi1988joint, rau1966detection}. The NL-material-based JTC architectures discussed here omit this step, potentially reducing latency and improving performance of such an optical processor. One of the classical correlator schemes discussed here is a four-wave mixing JTC \cite{pepper1978spatial, white1980real}. This architecture does not require EO-domain crossing in the Fourier plane and can be realized as a joint Fourier transform correlator (JFTC) architecture \cite{javidi1988joint}. Another subcategory of JTCs, featured in this paper, is non-linear JTCs (NL-JTCs) \cite{javidi1988joint, javidi1989nonlinear}. In these correlators, one introduces a nonlinear intensity threshold at the Fourier plane to control the output correlation contrast.

In this paper we identify the efficiency and scaling functions of the JTC. First we derive a minimum latency for convolution with respect to resolution. Next, we introduce and compare models for the intensity and signal-to-noise ratio (SNR) at the output of various JTCs. Finally, we explore the parameter space of the presented models through simulation of modulation depth, phase error, flatness,  tilt angle, and focal distance.

\section{Results and discussion}

\begin{figure}[t!]
\centering\includegraphics[width=8.7cm]{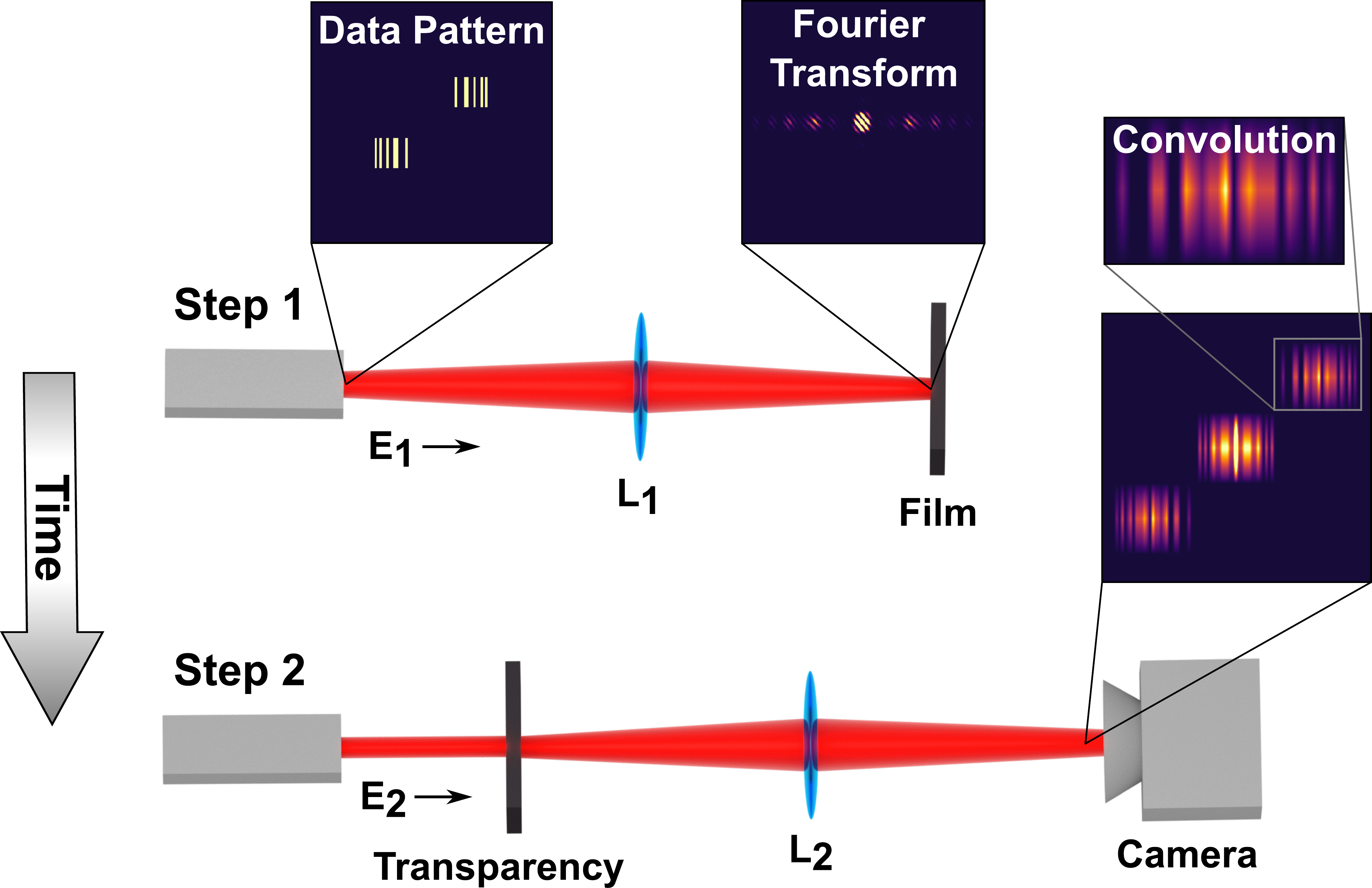}
\caption{The required computing overhead for convolution operation processing for machine learning (ML) tasks can be homomorphically accelerated using a Fourier-optics. The original joint transform correlator (JTC) is a two step process where the input data is projected on to field $E_1$ through lens $L_1$ to record the intensity pattern (a square of the Fourier transform) on to a piece of film, which is then developed into a transparency where a second field $E_2$ creates the inverse Fourier transform and the resulting three terms, including correlation or convolution, are recorded by a camera. Today this can be computer-controlled by replacing the film with a CCD and the transparency with an amplitude SLM.}
\label{fig:goodman}
\end{figure}

The idea of the classical JTC as a way to optically generate the full convolution (phase and amplitude), see Fig. \ref{fig:goodman}(a), was first developed by Goodman \cite{goodman2005introduction}. The original JTC was a two step process. In the first step, data sets $ h\left(x,y\right)$ and $g\left(x,y\right)$ are placed on a coherent beam $E_1$, which propagates through lens $L_1$. This lens acts as an approximate spatial Fourier transform and the result is recorded on film. In the second step, the film is developed into a transparency, capturing the Fourier transformed spatial data from the first step. This transparency is placed in the field of a new collimated beam $E_2$. The transparency acts to multiply the incoming beam by the intensity of the Fourier transform of the original spatial data. The intensity operator squares the spatial data in the Fourier plane. The second lens, $L_2$, performs an approximate Fourier transform on the squared spatial data resulting in an intensity pattern with the auto-convolution/auto-correlation of the entire input pattern, including both $h(x,y)$ and $g(x,y)$, at the CCD.

The output intensity contains four terms \cite{goodman2005introduction} (8-22)
\begin{align}
\label{eq:output_terms}
U\left(x,y\right) &= \frac{1}{{\lambda f}} \times \notag \\
&\times \big[ h({x},{y}) \otimes {h^*}( x, y) + g(x,y) \otimes g^* ( x, y) + \notag\\
& + \,\,h(x,y) \otimes g^*( x, y) \otimes \delta (x - R_x,y - R_y) + \notag\\
& + \,\,h^*( x, y) \otimes g(x,y) \otimes \delta (x + R_x,y + R_y)
 \big]
\end{align}
contains four distinctive terms: (1) the off-center terms $h \otimes g^*$ and $h^* \otimes g$, and (2) the central terms $hh^*$ and $gg^*$ along with a diffraction term from the portion of $E_2$ unmodulated by the transparency. This configuration can generate correlation of two supplied images $h$ and $g$, or convolution by mirror-reflecting one of the input images to get, e.g. $h^*$ and $g$ on the input. 

The original JTC is often performed by replacing the film with a CCD and the transparency with an amplitude SLM. The disadvantage of the original JTC is that there must be an optical to digital electronic stage between the CCD and the SLM which is a noisy and time/energy costly multi-step process.

The latency of all JTCs can be divided into two components: 1. the time of the light's path through the lens to generate each transform, and 2. time of the nonlinearity. The original JTC is the slowest because the nonlinearity consists of a discrete step between film and projection. While taking the most time, the scaling of the nonlinear step is constant, O(1), as the nonlinear step is completely parallel. Less obvious is the scaling of the light path through the lens. The JTC is dependent on the lens approximating the Fourier transform. This is only possible when the paraxial approximation is observed. The paraxial approximation is dependent on focal distance and image area where the maximum resolution elements $N_{max} = n^2$, where n is the edge size, are related to the focal distance by \cite{pepper1978spatial}: 
\begin{equation}
\label{eq:resolution_constraint}
N_{max} \approx 16f/\pi\lambda
\end{equation}
As the focal distance increases, so does the distance the light must follow. Using Eq. \ref{eq:resolution_constraint}, we derive the latency of the light path:
\begin{equation}
\begin{split}
\tau &= \frac{1}{c}\left(\sqrt{ \frac{\pi \lambda^2 N_{max}^{2/3}}{32} + \left(\frac{N_{max} \lambda \pi}{16} \right)^{2}} + \frac{N_{max} \lambda \pi}{16}\right) \\
&\Rightarrow \mathcal{O}(n^2)
\end{split}
\end{equation}
See Supplemental Material for derivation, Eq. S1-S12.


While the lens time is fixed, the original JTC can be modified to reduce the nonlinear time by replacing the film and transparency with an intensity dependent nonlinear process such as amplification, nonlinear absorption,  including carrier generated absorption, two-photon absorption, saturable absorption, or complex $\chi^{(3)}$ absorption; or harmonic generation. For an optical field carrying $n$ distinct signals incident on a non-linear material, the matter response can be described in the framework of the non-linear optics as a generalized polarization tensor
 \begin{equation}
    P^{(n)}(t) = \epsilon_0 \chi^{(n)} E^n(t) + c.c.
    \label{eqn:polarization_general}
 \end{equation}
 where $\epsilon_0$ is the vacuum permittivity and $\omega_i$ are the optical frequencies contributing into the input field. Hence, Eq. \eqref{eqn:polarization_general} contains a polynomial of a degree $n$ with all possible combinations of the constituent signals contributing to the input field. The term $c.c.$ introduced in Eq.\eqref{eqn:polarization_general} signifies the complex conjugate and will be omitted in the further discussion for brevity.

\begin{figure}[b!]
\centering\includegraphics[width=8cm]{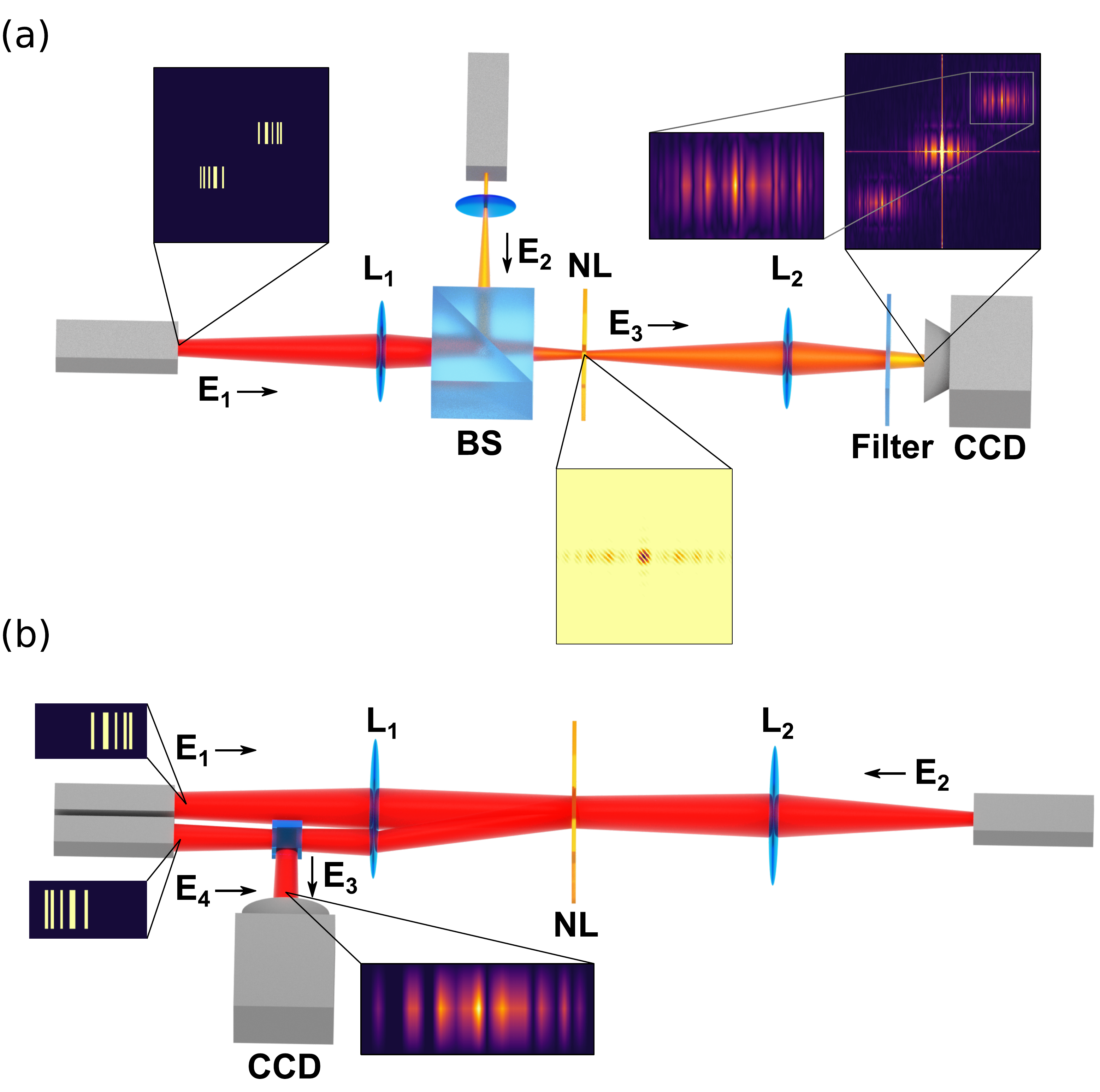}
\caption{Nonlinear JTC configurations: (a) a pump-probe JTC where the auto-correlation of the pump $E_1$ is formed on the probe $E_2$, by an amplitude change proportional to intensity forming the square of the electric field to create the field $E_3$ at the CCD and harmonic generation. The input and output signals are separated by a wavelength filter; (b) the optical holographic 4f JTC where the fields $E_1$, $E_4$, and $E_2$, are mixed in the nonlinear material to form the correlation by optical phases conjugation in field $E_3$ at the charge couple device.}
\label{fig:configurations_direct}
\end{figure}

Unlike the classical JTC, in the absorption-based nonlinear JTC the two fields $E_1$ and $E_2$ are acting simultaneously being projected by lenses $L_1$ and $L_2$ correspondingly through the beamsplitter onto a NL material placed in the lenses' mutual focal plane, see Fig. \ref{fig:configurations_direct}(a). The field $E_1$, being loaded with the input data, generates spatial variations of the local absorption coefficient in the NL material driven by the third order non-linearity effects. This process can be described by Eq. \eqref{eqn:polarization_general} with $n=3$ such as
\begin{equation}
    P^{(3)}(t) = \epsilon_0 \chi^{(3)} |E_1(t)|^2 \cdot E_2(t)
\end{equation}
The second field $E_2$, acting as a probe, propagates through the NL material being amplitude modulated by these absorption coefficient variations proportional to the squared signal imprinted in the pump field $E_1$. The second spatial Fourier transform by lens $L_2$ generates an autocorrelation of the signal in $E_1$ due to the equivalence of an auto-correlation and squaring in the Fourier domain. Finally, the pump field is spectrally filtered. In Fig.  \ref{fig:configurations_direct}(a) we show one of the possible realizations of this JTC.

To quantify the signal portion of the output intensity, we start with the output intensity $ I\left(u,v\right)$ at the CCD, and work back through the Fourier transform of $L_2$. 
Due to the linearity of the Fourier transform, we can separate these two components into a direct current (DC) term $U_{DC}$, and modulated term $U_{mod}$:
\begin{equation}
I\left(u,v\right) = \sqrt{\frac{\epsilon_0}{\mu_0}} \frac{A_{pr}^{2}} {\lambda^2 f^{2}} \lvert\mathscr{F}\left\{U_{DC}\right\}+\mathscr{F}\left\{\mathrm{U_{mod}}\right\}\rvert^{2} \label{eq:intensity}
\end{equation}
where $A_{pr}$ is the probe amplitude. The amount of power that is placed either on $U_{DC}$ or $U_{mod}$ is determined by the nonlinearity of the material and the distribution of pump intensity on the nonlinear material. The strength of the modulated portion can be approximated by the dynamic range of modulated probe field, $ U_{mod} $, found by taking the difference between the modulation of the strongest and weakest frequency components of the pump field, $\zeta_{max}$ and $\zeta_{min}$, in the Fourier plane:

\begin{equation}
U_{mod} \approx \exp{\left(\alpha\left|{\zeta_{min}}\right|^2\right)}-\exp{\left(\alpha\left|{\zeta_{max}}\right|^2\right)}\label{eq:modulationdepth}
\end{equation}
where $A_{pu}$ is the pump amplitude, and $\alpha=-\;\beta z_0A_{pu}^2\; / \lambda^2f^2$. By approximating modulation depth by the first term of a Taylor expansion of Eq. \ref{eq:modulationdepth}, assuming  $\alpha_0 \approx 0 $ and positive real values for $\chi^{(3)}$ and amplitudes, we obtain an approximation for the intensity of the modulated component at the output:

\begin{equation}
\label{eq:absorption_JTC_efficiency}
\begin{split}
I_{mod} (u,v)\approx  \Phi^2(\lambda, f) \left|\chi^{\left(3\right)}\right|^2\ \left|\mathscr{F}{(| \zeta_{max} |^2-|\zeta_{min} |^2 )}\right|^2
\end{split}
\end{equation}
where $\Phi(\lambda, f)$ is a coefficient that scales as $f^{-3}$ and $\lambda^{-4}$. The significant part of the probe field does not get modulated by the pump and is transformed into a $\text{sinc}$-function at the center of the output. These overlays the autoconvolution terms, $h \otimes h^{*} $ and $g \otimes g^{*} $ from Eq. \ref{eq:output_terms}. Hence, the convolution or correlation can't be easily separated from the output intensity as the corresponding terms depend on the input data $h$ and $g$. However, for sufficiently long sequences, the average efficiency of either the correlation or convolution we are interested in becomes Eq. \eqref{eq:absorption_JTC_efficiency} divided by a factor of four, see Supplemental Material, Fig. S2-S7.

\begin{figure*}[t!]
\centering\includegraphics[width=18cm]{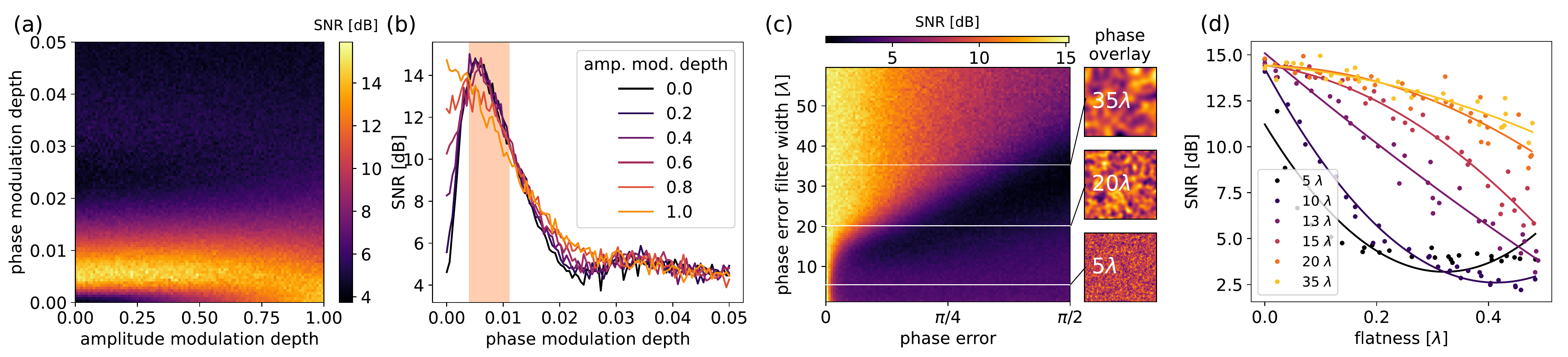}
\caption{Simulation results showing: (a) the SNR of the optical convolution vs the numerical computation of the amplitude JTC, with phase and amplitude modulation; (b) a small region near phase modulation $\approx$ 0.008 where phase modulation has the same effect as amplitude modulation; (c) SNR swept against phase error filter width and phase error amplitude std. dev. with insets showing the spatially filtered phase noise at $5\,\lambda$, $20\,\lambda$, $35\,\lambda$ spatial filter widths; (d) SNR versus flatness where flatness is calculated from a simulated interferogram for each point in (c) and plotted for varying phase error filter widths from 5$\lambda$ to 35\,$\lambda$. It shows saturating SNR for filter widths greater than 20\,$\lambda$ $\approx2\times$ the features size, 8.4\,$\lambda$.}
\label{fig:amplitude_quad}
\end{figure*}

It is important to note that realizable NL materials always have both nonlinear index, $ n_2 $, and nonlinear absorption $ \beta $. Strictly speaking, the probe beam is modulated in both amplitude and phase, $ E = A \exp\left[(-\beta + j n_2) I d\right] $ instead of the amplitude-only modulation assumed above. However, in the limit of small phase values, the exponential phase term
\begin{equation}
\begin{split}
|\exp\left(j n_2 I d\right)|^2 \sim \left( 1 + \left(n_2 I d\right)^2\right)
\label{eq:phase_modulation_results_in_amplitude}
\end{split}
\end{equation}
translates into an amplitude modulation. This result is observed in simulation of the absorption JTC for small phase modulations [bottom left region Fig. \ref{fig:amplitude_quad}(c)].

In a four-wave mixing configuration, an optically active material with a third-order non-linearity is placed in the 2f-focal point of a 4f-filter, see Fig. \ref{fig:configurations_direct}(b). Three monochromatic laser beams enter the system, with $E_1$ and $E_2$ being amplitude modulated with distributions $u_1(x,y)$ and $u_2(x,y)$, and $E_4$ acting as a probe field with the distribution $u_4(x,y)$. Following the formalism outlined in \cite{pepper1978spatial}, specifically, see eq-s (10-12), and assuming the approximation of stationary fields, and that $\Delta x, \Delta y >> \Delta z$, where ${\Delta x,\Delta y,\Delta z}$ are the dimensions of the non-linear material, one arrives at the following expression for the intensity at the CCD.
\begin{equation}
I_3\ (x,y,0)= \sqrt{\frac{\epsilon_0}{\mu_0}} \frac{\pi^2 \omega^2 z_0^2} {16c^2 \lambda^8 f^8} |\chi^{(3)}|^2 |U|^2
\label{eq:four_wave_intensity}
\end{equation}
One can see that $ I_3 $ is directly proportional to the input fields intensity and the magnitude square of the third-order nonlinear susceptibility $ \chi^{\left(3\right)} $ which is substantially easier to analyse as compared to the absorption NL-JTC discussed above, see Eq. \eqref{eq:absorption_JTC_efficiency}.

\begin{figure}[b!]
\centering\includegraphics[width=8.5cm]{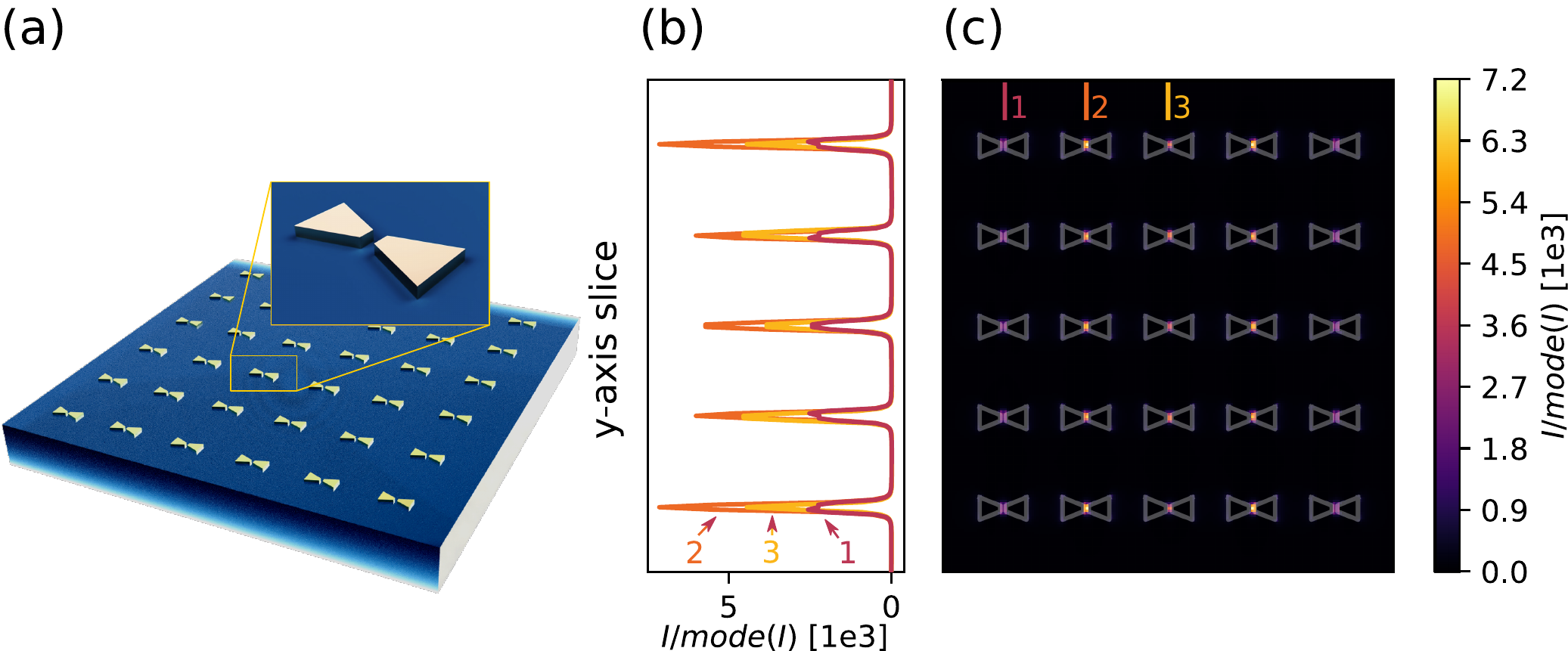}
\caption{ (a) Metamaterial structures such as gold bow-tie antennas (b) at plasmonic resonance on an SiO$_{2}$ substrate simulated with finite difference time domain (FDTD) using MEEP \cite{oskooi2010meep} (c) enhance the electric field, with maximum intensity enhancement $>5\times10^{3}$ (b), increasing the  effect of the nonlinear material.}
\label{fig:metamaterials}
\end{figure}

In both the mixing and the JTC approaches, discussed above, the nonlinear process may be enhanced through metamaterial fabrication by a combination of patterning metallic nanoantennas \cite{fromm2004gap, muehlschlegel2005resonant} to create plasmonic resonators, see Fig. \ref{fig:metamaterials}(a), patterning high dielectrics to form anapoles \cite{grinblat2020efficient, gupta2020toroidal}, see Fig. \ref{fig:metamaterials}(b), and operating near the ENZ wavelength of an ENZ material such as ITO \cite{luk2015enhanced, capretti2015enhanced, alam2016large}, AZO \cite{carnemolla2018degenerate, caspani2016enhanced}, or an ENZ metamaterial  \cite{alam2018large, neira2015eliminating, kaipurath2016optically}. In all such structures the electric field is either confined to a smaller area or held in resonance for a longer time, enhancing the electric field locally and leading to a stronger nonlinear response..

To arrive at a value for the efficiency of the convolution we consider the SNR of a pixel in the CCD \cite{o1992charge}:
\begin{equation}
\label{eq:ccd_snr}
     SNR_{pixel} = \frac{\eta\Phi_p\tau}{\sqrt{\eta\Phi_p\tau + I_{dark}\tau + N_r^2  }},
\end{equation}
where the parameter $N_r =  \sqrt{N_t^2 + N_o^2}$ is defined over readout noise $N_r^2$, which is composed of the transfer noise $N_t^2$ and the output noise $N_o^2$; $\eta$ is the quantum efficiency, $\Phi_p$ is the photon flux, $\tau$ is the integration time, $I_{dark}$ is the dark current.

There is a relationship between the SNR required to observe N bits in the ideal ADC \cite{kester2005data}: 
\begin{equation}
\label{eq:snr_bits}
SNR_{dB} = 6.02 N + 1.76 dB
\end{equation}

Thus for 4, 8 and 16 bits at the ADC we require an SNR of 25.84 dB, 49.9 dB, and 98.08 dB respectively, demonstrating the high cost for multiple bytes of bit resolution.

The output amplitude of the four-wave mixing correlator is proportional to the input amplitude. While the spatial distribution of the field $U$ in Eq. \eqref{eq:four_wave_intensity} is dependent on the input data, the field is separable due to associativity of convolution/correlation operation

The output intensity is divided among the pixels of the CCD. The maximum resolution is bound by the paraxial approximation. For the four-wave mixing correlator this resolution can be found to be\cite{pepper1978spatial}: $N_{max} \approx 16f/\pi\lambda$, See Supplemental Material for derivation, Eq. S1-S8. Here, $N_{max}$ is the upper limit on the number of pixels in the output image. It may be possible to relax the paraxial approximation bound on resolution at the cost of decreased spatial SNR in the resulting convolution.

Then, in the four-wave mixing case, we arrive at an approximation for the average photon flux per pixel at the CCD:
\begin{equation}
\label{eq:four_wave_photon_flux}
\Phi_{p}\ \approx \frac{\pi^2 \omega^2 z_0^2} {16c^2 \lambda^8 f^8} |\chi^{(3)}|^2 I_{input} \frac{\lambda}{hc} L_p^2
\end{equation}
Here, $L_p^2$ is the area of the pixel. Combining Eq. \eqref{eq:four_wave_photon_flux}, Eq. \eqref{eq:snr_bits}, and Eq. \eqref{eq:ccd_snr} we now have a model of the four-wave mixing correlator that relates required input intensity and material $\chi^{(3)}$ to achieve an average pixel ENOB at the output CCD.

Comparing the result to the intensity of the four-wave mixing approach, Eq. \eqref{eq:four_wave_intensity} with intensity in the nonlinear JTC, Eq. \eqref{eq:absorption_JTC_efficiency}, we see that while both approaches scale in terms of wavelength with $ \lambda^{-8} $, the absorption JTC approach scales with focal distance $ f^{-6} $ versus $ f^{-8} $ in four-wave mixing. This can be explained by the four-wave mixing approach using four fields, each obeying the inverse square law, versus the absorption configuration using three fields, two patterns in the pump and a probe, each with the inverse square law. Therefore, ignoring other parameters, the absorption approach will be more efficient with a square of the focal distance.

To simulate the convolutions resulting from amplitude JTC and 4-wave mixing correlator we created a Python model using Fresnel transfer function propagation theory \cite{voelz2011computational}. Here, propagation is modeled with the Fresnel approximation as a convolution with the impulse response:
\begin{equation}
h\left(x,y\right) = \frac{e^{jkz}}{j \lambda z} \exp\left[\frac{jk}{2z} \left(x^2 + y^2\right)\right]
\label{eq:Fresnel_approximation}
\end{equation}
Each lens is modeled as a simple phase transform: $ \exp\left(\left(x^2 + y^2\right) jk/\left(2f\right) \right) $ where $ f $ is the focal distance. For the detailed algorithm and the table of parameters we refer the reader to Supplemental Material Sec.3.

The output of the simulation is compared with an exact convolution of the inputs to generate an SNR along each pixel of the simulation as a metric to evaluate the result.

SNR in the simulation is defined using the image mean-squared sense \cite{gonzalez2008digital}:
\begin{equation}
SNR_{image} = 10 \log_{10} \left[ \frac{\sum_{x=0}^{(n_x-1)} r\left(x\right)^2 }{\sum_{x=0}^{(n_x-1)} \left[ r\left(x\right) - t\left(x\right)\right]^2 } \right]
\end{equation}
Where $r$ is the reference image, the result of the mathematical convolution,  and $t$ is the test image, the result of the optical simulation. Each pixel in the calculation is a pixel of the simulation grid which is fixed for all simulations at $n_x$ = 32768. However, in the case of the amplitude JTC, the convolution is only a portion of the result, Fig. \ref{fig:configurations_direct}, in this case the simulation result was cropped to  $n_x$ = 3199.

A sweep of phase modulation depth vs amplitude modulation depth, Fig. \ref{fig:amplitude_quad}(a), shows that small modulation of the phase in terms of intensity results in an equivalence to amplitude modulation, see Eq. \ref{eq:phase_modulation_results_in_amplitude}. In a region around phase modulation $\approx 0.008$, the SNR of the  convolution actually increases with decreasing amplitude modulation, Fig. \ref{fig:amplitude_quad}. While theoretically interesting, it is likely impractical to achieve such small phase modulations consistently above the system phase noise floor.

Phase error in coherent optical systems is a noise signal varying in time and/or space, applied to the phase component of the optical field (i.e. $\phi$ in $A exp\left(\omega t + \phi \right)$). Time varying phase noise is due to the phase noise in the laser source, vibration of optical components, and air turbulence. Spatial phase noise is due to variations in the surface of optical components. Lenses are specified in terms of flatness, which is often defined as surface variance at the operating wavelength as measured by an interferogram. This definition, however, does not specify the spatial frequency of the variance. 

Phase error is simulated by applying a Gaussian noise filtered by a spatial Gaussian filter of a specified spatial width in $\lambda$, normalized by standard deviation, and multiplied by the desired amplitude. The phase error is applied at nonlinear plane. The phase noise was swept from $0$ to $\pi/2$, the results show a degradation with increasing phase error, Fig. \ref{fig:amplitude_quad}(c). The spatial frequency of the noise, determined by the Gaussian filter width, has a dramatic effect on the degradation of the SNR with respect to the strength of the phase error, Fig. \ref{fig:amplitude_quad}(c) for $5\lambda, 20\lambda, \text{\;and\;} 35\lambda$.

Additionally, flatness in terms of $\lambda$ was calculated by a simulated interferogram generated for each point of the  Fig. \ref{fig:amplitude_quad}(d). This metric allows an SNR to be approximated given the flatness specified by the optical components of the system.

Tilt is simulated by applying a phase term \cite{voelz2011computational} across the $x$-axis of the wavefront to describe the difference in phase terms in the angled configuration of the amplitude JTC [Fig. \ref{fig:configurations_direct}(c)]:

\begin{equation}
u_{out}(\alpha) = u_{in} \exp\left[jk \tan \left(\alpha\right)\right]
\end{equation}

\begin{figure}[b!]
\centering\includegraphics[width=8.7cm]{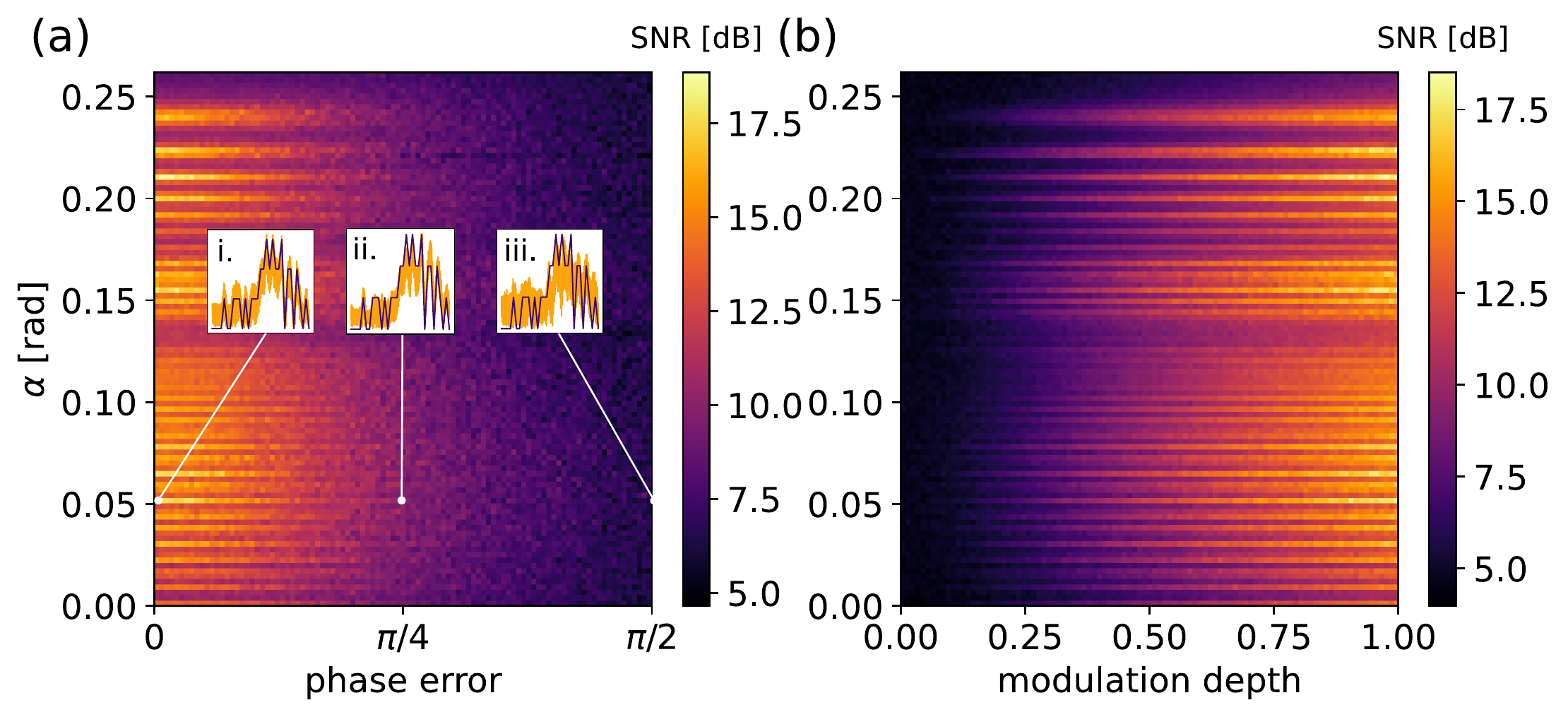}
\caption{Angle $ \alpha $ between the pump and probe in the amplitude JTC simulation shows a sinusoidally varying angle dependent error (a) that increases in peak amplitude as the angle $ \alpha $ grows. The angle induced phase error appears independent of the additional swept phase error ((a) x-axis) and the swept modulation depth (b).}
\label{fig:pump_probe_angle}
\end{figure}

The pump-probe angle was swept between $0^{\circ}$ and $15^{\circ}$ against a modulation depth swept from 0 to 1, Fig. \ref{fig:pump_probe_angle}. The resulting SNR has regions of lower signal that are a function of the tilt angle, $ \alpha $. 

\begin{figure*}[t!]
\centering\includegraphics[width=18cm]{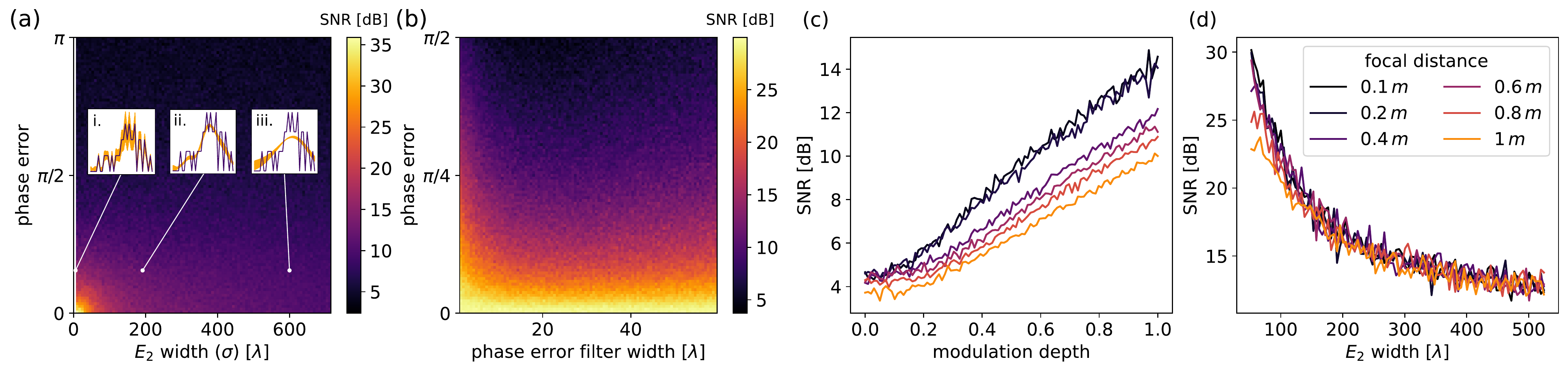}
\caption{Four-wave mixing JTC simulation with SNR swept against the phase error and carrier beam $E_2$ width: (a) increasing SNR with phase error and with the width of the carrier beam $E_2$ which has the effect of a Gaussian blur on the output convolution; (b) SNR swept against phase error and phase error filter width at constant $E_2$ width of $\sigma = 1\,\lambda$ which shows decreasing SNR with increasing phase error and insignificant dependence on phase error filter width; (c) SNR swept against modulation depth for the amplitude JTC which shows little to no dependence on focal distance while simulation of the four-wave mixing against $E_2$ width; (d) insignificant dependence on focal distance except when $E_2$ is $< 100 \lambda$, when simulated with $M = 2^{18}$ and 2000 pixel pitch.}
\label{fig:four_wave_mixing_result}
\end{figure*}

The four-wave mixing correlator, Fig. \ref{fig:configurations_direct}, is simulated in a similar manner to the amplitude JTC, however instead of applying an intensity derived modulation, the mixing result of Eq.(11) in \cite{pepper1978spatial} is generated in the Fourier domain and then propagated through the final lens. The simulation was swept over phase error and $E_{2}$ carrier width, Fig. \ref{fig:four_wave_mixing_result}(a), to identify tolerance to phase variation and beam width. Additionally, phase error was swept against the phase error filter width in units of $\lambda$ to identify tolerance to varying spatial phase noise,  Fig. \ref{fig:four_wave_mixing_result}(b).

As the focal distance becomes larger the 4f system approaches the Fourier transform due to the paraxial approximation in the derivation of Eq. \eqref{eq:Fresnel_approximation}. Through the use of Eq. \eqref{eq:Fresnel_approximation} the simulation assumes the paraxial approximation, and small focal distances will result in an error that is not simulated \cite{goodman2005introduction, voelz2011computational}. However, for larger focal distance where the paraxial approximation applies the effect of focal distance over modulation depth is limited, Fig. \ref{fig:four_wave_mixing_result}(c). 

In the case of the four-wave mixing correlator, focal distance has an effect when the carrier beam, $E_2$, has a small spatial extent, Fig. \ref{fig:four_wave_mixing_result}(d). As the size of the carrier beam increases $E_2$ the SNR of the result decreases. This is mostly independent of focal distance unless  $E_2$ is small ($< 100 \lambda$).

\section{Conclusion}
Nonlinear materials placed within a 4f system provide an effective mechanism for full optical convolution. As convolution becomes a greater part of the computing landscape with the continued increase in neural network applications, free-space optical convolution is increasingly appealing due to its high spatial bandwidth and low latency. The efficiency of the approach is dependent on nonlinear optical characteristics of the materials. These properties continue to be enhanced with exploration of ENZ systems and field confining structures.

In the future, the free-space optical components can be replaced by the on-chip lenses which have been achieved in the nanophotonic convolver \cite{liao2020ai}, on-chip deep learning systems \cite{fu2021chip,chen2021chip,zarei2020integrated,wang2019chip} and integrated tunable varifocal lenses performing like error backpropagation in neural networks \cite{zarei2021inverse}. Once combined with on-chip lenses, our nonlinear JTC will have high energy efficiency, compact volume and high speed, and therefore promotes the potential convolution-related applications in many aspects, including image classification with large and deep convolutional neural networks, speech recognition and translation with the combination of convolutional neural networks and deep recurrent neural networks, autonomous driving by mapping raw pixels with convolutional neural networks, inverse design (like nanophotonic / plasmonic structures, self-adaptive microwave cloak) problem solving based on convolutional neural networks, and robotic manipulation with deep reinforcement learning.
\vfill

\end{document}